\documentstyle[aps,multicol,epsf]{revtex}
\begin{document}
\draft

\title{Worm Structure in Modified Swift-Hohenberg Equation for Electroconvection}

\author{Yuhai Tu}

\address{IBM Research Division, T.J.
Watson Research Center, P.O. Box 218, Yorktown Heights, NY 10598}
\date{\today}
\maketitle

\begin{abstract}
A theoretical model for studying pattern formation in electroconvection
is proposed in the form of a modified Swift-Hohenberg equation. A localized
state is found in two dimension, in agreement with the experimentally 
observed ``worm" state. The corresponding one dimensional model is also 
studied, and a novel stationary localized state due to nonadiabatic effect 
is found. The existence of the 1D localized state is shown to be responsible 
for the formation of the two dimensional ``worm" state in our model.
\end{abstract}

\begin{multicols}{2}
\narrowtext
When a system is driven away from its equilibrium state,
it often responds in forming regular patterns. 
Pattern formation phenomena is rather ubiquitous, occurring in many different 
physical, chemical and biological systems. One of the most carefully studied 
systems is the Rayleigh-Benard convection (RBC), where high precision
experimental measurements can be carried out and compared quantitatively 
with theoretical studies. The careful study of this rich physical system for 
the past twenty years 
has greatly advanced our understanding of 
systems far away from equilibrium\cite{Cross}.

The study of localized structure in nonequilibrium system has received a 
great deal 
of attention since being observed experimentally in binary-mixture 
RBC\cite{Moses}. Even though binary-mixture 
RBC is a highly dissipative
system, the localized structures behave much like soliton in integrable
systems. For example, in certain parameter range, they can pass through
each other without changing their structures. 
On the theory side,
Thual and Fauve\cite{Thual} were the first to study the behavior of a 
subcritical
complex Ginzburg-Landau equation and found that in certain parameter range,
there are indeed localized pulse solutions. 
The basic ingredients for the existence 
of localized structure are: 1)there has to be
linear bistability, which guarantees the local stability of the
peak and the tail of the pulse; 2)the nonlinear dispersion
(the complex part of the coefficients for the nonlinear terms) is needed to 
stabilize
the front connecting the peak and the tail of the localized solution. 
Both of these requirements seem to be consistent with the experimental
system, because the initial bifurcation in binary-mixture RBC is 
subcritical Hopf bifurcation.
Much work has since been devoted along these lines to understand the details
of the experimental results\cite{riecke,brand1}. 

Most of the experimental results in binary-mixture RBC were obtained in quasi one
dimension, i.e., in a thin annulus\cite{kolodner}. Further efforts to extend these findings
to two dimensional system have not revealed any similar 2D localized 
state as 
in 1D except for some time dependent patchy structure\cite{steinberg} and some long time transients\cite{boden}. 
Recently, M. Denning et al\cite{dennin} studied electroconvection in 
nematic
liquid crystal very carefully. Due to the anisotropic nature of the liquid 
crystal, they
found that the initial instability of the system is towards forming oblique
rolls with certain angle with respect to the director of the nematic
liquid crystal. They also found that the initial bifurcation is Hopf 
bifurcation\cite{dennin2}.
Depending on the electrical conductivity, the pattern they observed 
above onset is either spatially extended time dependent state/spatial-temporal chaos
state(STC) or some isolated localized state, which they named the ``worm"
state. The worm state is localized in the direction perpendicular to the
director of the liquid crystal, but is extensive in the parallel direction.
The worm state can move in the parallel direction. The internal structure
of the worm state seems to consist of both orientations of the oblique rolls, and internal 
roll structure is moving relative to the motion of its envelop.
It is the goal of this paper to understand the interesting 
structure of the worm state.

There are a few
well established methods for studying pattern formation. Notably among them is 
the amplitude equation
formalism which was used in [3] and related works\cite{brand1,riecke} 
to study the pulse pattern
in binary-mixture RBC. Amplitude equation describes the large scale 
and long time behavior of the
envelop of the pattern. It is perturbative in nature
and describes the system accurately at parameters close to the onset.
The shortcoming of the amplitude equation is thus quite obvious.
In writing down the amplitude equation, one has already broken the
full spatial symmetry of the original system which might be important in 2D. 
In the case of 
subcritical bifurcation, the amplitude equation can only give qualitative
results because the amplitude does not scale with the small parameter(reduced
Rayleigh number). 
For the problem at hand, there are even more severe limitation for the 
amplitude equation. Since the spatial extension of the worm state in the
perpendicular direction is only about a couple of wavelengths, there is no
seperation of length scales to justify the use of the amplitude equation. 

To understand the experiments
qualitatively, a phenomenological model is often useful. 
The Swift-Hohenberg(SH) equation\cite{sh}
is an order parameter equation with the full symmetry of the original problem,
and its linear properties agree with that of the original problem. The SH equation is 
phenomenological in nature, and usually can not be derived from the original
equations. 
For different experimental systems, 
The Swift-Hohenberg equation can be modified in different ways 
from its original form to accommodate different physical situations and
symmetry requirements, e.g., the Non-Bousinesq
effect, mean flow effect\cite{meanflow} 
,Hopf bifurcation\cite{csh,besthorn} and rotating convection\cite{ctm}. Because of its 
versatility and simplicity, the modified Swift-Hohenberg equation(MSHE) has 
become instrumental in understanding many pattern forming systems. 

In writing down the modified Swift-Hohenberg equation for 
electroconvection, we know that the equation has to be anisotropic even at
the linear level, and the equation has to be complex because the initial
bifurcation is Hopf bifurcation. Let $\phi(\vec{x},t)$ be the complex
order parameter, we can write the order parameter equation as:
\begin{eqnarray}
\partial\phi/\partial t&=&(\epsilon+i\omega)\phi-\sigma ((\partial_x^2+q_x^2)^2+
b(\partial_x^2+q_x^2)(\partial_y^2+q_y^2)\nonumber\\
&+&(\partial_y^2+q_y^2)^2)\phi+iv_g ((\partial_x^2+q_x^2)+a(\partial_y^2+q_y^2))\phi\nonumber\\
&+&g_0 |\phi|^2\phi +g_1 |\phi|^4\phi
\end{eqnarray} 
$\epsilon$ is the reduced Rayleigh number, $\omega$ is the Hopf
frequency, $\vec{q}=(q_x,$ $q_y)=(cos\theta,$ $sin\theta)$ is the linearly
most unstable wavevector, b is an anisotropic parameter with the constrain
$|b|\le 2 $, and $\sigma$ is a complex constant. The first two lines on the RHS of 
eq. (1) represent the linear properties of the elctroconvection system. 
It is easy to see that the system is most unstable at $|k_x|=q_x$ and 
$|k_y|=q_y$ for $\phi\sim\exp (ik_x x+ik_y y)$.
$v_g$ is proportional
to the group velocity and $a$ is another anisotropic parameter. When $a=1$, the group velocity is along the wavevector direction $\hat{q}$. The last
line on the RHS of eq. (1) contains the nonlinear coupling terms with complex 
coefficients $g_0$ and $g_1$. In principle, the nonlinear terms can also be
anisotropic, we only include the simplest terms possible here. 

Since it is the goal of this paper to find the localized worm state,
we focus our attention on the subcritical case where $\epsilon<0$, 
$Re(g_0)>0$ and $Re (g_1)<0$. 
We can easily get rid of the $i\omega$ term in the linear part of the equation
by a change of variable $\phi=e^{i\omega t}\phi$, so we will set $\omega=0$ for
now on. There are five real parameters: $\epsilon$, $a$, $b$, $\theta$ and
$v_g$ and three complex parameters: $\sigma$, $g_0$ and $g_1$ for this model.
We have numerically studied the MSHE extensively
in the parameter space and identified certain parameter region where
localized worm state is observed.

To demonstrate the existence of the worm state, 
we first show the
behavior of the eq. (1) for a particular set
of parameters: $\epsilon=-0.2$, $a=1$, $b=0$, $\theta=23^o$, $v_g=0.5$,
$\sigma=1.5$, $g_0=3+i$ and $g_1=-2.75+i$. The equation is simulated in
systems of size $64 \times 64$, $128\times 64$ and $256\times 64$ with periodic
boundary condition using both the second order finite difference method and 
spectral method with discretization $\Delta x=\Delta y=0.5,1.0$ and
time step $\Delta t=0.001,0.01$. 
We start the system with random initial condition 
with large enough amplitude. The system quickly
organizes itself into the worm like state. A snap shot of the 2D pattern for $Re(\phi(x,y))$ after the initial 
transient is shown in figure 1a.

A slice of the 2D pattern along the y direction at $x=45$ 
is shown in figure 1b. The localization of the worm states in the y direction
is apparent from fig. 1b. The Worm states travel in the x direction. 
From their length, 
the worm states in our simulation can be divided into two categories, which 
we call
long worm and short worm. The length of the short worm does not change with 
time, and is usually $\sim 20$, which is 3 basic wavelengths long. They 
travel in the x direction with constant velocity proportional to $v_g$. 
An example of a short worm can be seen near the bottom of fig.1a. The
long worm's length grows with time and eventually extends over the whole length
of the system because of the periodic boundary condition.

We have tested the sensitivity of the worm pattern to the parameters in
our model. We find that there is a finite range of parameters where the worms
appear. For example, if we change the value of $\epsilon$ while keeping the
rest of the parameter unchanged, worm exists for $-0.10>\epsilon>-0.25$. When
$\epsilon$ is too small, there is no pattern; and when $\epsilon$ is too
big, the pattern becomes extended instead. The worm state is quite insensitive 
to the
values of $a$ and $b$, as long as $a\sim 1$ and $|b|<2$. For $b=2$ and $a=1$,
the model becomes isotropic and the worm structure gives away to
time dependent
patchy structure\cite{besthorn}. 
The velocity $v_g$ is important to give the worm a group
velocity. The wavevector angle $\theta$ has to be small enough $\theta\le
35^o$ to make the worm perfectly aligned in the x direction. 
There are
also finite regions in the parameters $g_0$, $g_1$, $\sigma$ where worm states 
are observed.

The worm states interact strongly with each other. 
When two short worms collide, they
come out of the collision without changing their characteristics.
When a short worm collides with a long worm, the short worm sometimes
disappears. When two long worms approach each other off center, oblique rolls
are excited in the region of their overlap until the worms pass through each other or one of the worms disappears.
When two long worms collide head on, they stop each other and form a
well defined boundary between them.

For the short worm, because of the spatial extention in both directions
are about the same order, the formation of the short worm should be due 
to strong interaction between
the two dimensions. However, for the long worms, due to the extensiveness 
of the worm in the x direction, we are able to separate the dependence
in the two dimensions and therefore gain more understanding of the mechanism
for the localization in the y direction. Indeed, Fourier analysis of the long worm along x direction shows that it is a good approximation to assume the x 
dependence to be a simple plane wave:
\begin{equation}
\phi(x,y,t)=\psi(y,t)\exp (ik_x x)
\end{equation}
If we substitute the above ansatz into the original equation (1), we obtain
a one dimensional dynamical equation for $\psi(y,t)$. For simplicity, we
set $a=1$ and $b=0$:
\begin{eqnarray}
\partial\psi/\partial t &=&(\tilde{\epsilon}+i\tilde{\omega})\psi-\sigma(
\partial_y^2 +q_y^2)^2\psi+iv_g(\partial_y^2+q_y^2)\psi\nonumber\\ &+&g_0 |\psi|^2\psi+g_1 |\psi|^4
\psi
\end{eqnarray}
where, $\tilde{\epsilon}=\epsilon-Re(\sigma)(q_x^2-k_x^2)^2$ and
$\tilde{\omega}=\omega-Im(\sigma)(q_x^2-k_x^2)^2+v_g(q_x^2-k_x^2)$. For the
same reason as in two dimension, we can set $\tilde{\omega}=0$.

We have studied the above 1D MSHE carefully. The numerical scheme 
is the same as in the two dimensional case, and we also started with a
random initial condition with sufficient amplitude. In order to compare it to 
the two dimensional case, we have set the parameters $\sigma=1.5$,
$v_g=0.5$, $g_0=3.0+i$ and $g_1=-2.75+i$ to be the same as in the 2d 
calculation. We can vary $\tilde{\epsilon}$ because the value of $k_x$ is 
undetermined a priori. For $q_y=sin(\theta)$ with $\theta=23^o$,
we found a finite range of $\tilde{\epsilon}$ values, where localized structure is
observed $-0.15>\tilde{\epsilon}>-0.5$. A space-time plot of $Re(\psi(y,t))$ 
is shown in figure 2. It is clear that the
system evolves to a final state with two localized pulses. Most
remarkably, the pulses are not moving even in 
the presence of the group velocity term in eq. (3). 

We found that the pulse solution can be written as:
\begin{equation}
\psi(y,t)=A(y)\exp (i\alpha (y,t))
\end{equation}
where the amplitude $A(y)$ is independent of time and is localized with
a width of $\sim 10$. Shifting the peak position to be at $y=0$, the shape of 
the pulse is symmetric around y=0: $A(y)=A(-y)$
. The phase of the pulse depends on time linearly:
\begin{equation}
\alpha(y,t)=\alpha_0 (y)+\Omega t
\end{equation}
with $\Omega=-.25$. The shape of the time independent phase $\alpha_0(y)$
is depicted in figure 3(b). From figure 3(b), we see that the phase is symmetric
around $y=0$: $\alpha_0(y)=\alpha_0(-y)$. Away from the peak of
the pulse, the phase is roughly linear in y:
$$
\alpha_0 (y)\sim k_y |y| +const \;\;\;\; |y|>5
$$
with $k_y\sim 1$.

As we pointed out earlier in our paper, the existence of localized state 
in subcritical equation with complex coefficient was known\cite{Thual}.
However, a stationary localized pulse in the full equation including
group velocity term is observed for the first time here. If one were able
to eliminate the small scale structure and write the full equation in terms
of the amplitude equation, one could use two amplitudes characterizing
the left and the right moving wave packets. As shown in the work of Brand and
Diessler\cite{brand2}, the left and the right moving pulses often can pass through each other
without altering their own characteristics. Upon tuning the inter-coupling 
between the left and the right moving pulses, the pulses can form bound state
which do not move in either directions. However, the structure of the bound 
state is such that the amplitudes of both the left and the right moving 
pulses are strongly suppressed in their coexisting region. In the middle of 
the  bound state, both amplitudes are very small, and overall, within the 
bound state, both pulses still keep their own identities.

The pulse structure we observed here can not be explained as the bound state
of the left and the right moving pulses in the amplitude equation 
because the localized
state here does not have local minimum at the center. On the contrary, the 
amplitude is the maximum at the center of the localized pulse. Furthermore, 
in our simulation, we
never observe any individual traveling pulse and the stationary localized structure 
always forms spontaneously as an whole object. The two
halves of the pulse seem to have opposite phase velocity 
$v_p=\pm \Omega/k_y$. However, the periodicity in the phase $2\pi/k_y$ 
is much smaller than the linearly
most unstable wavelength $2\pi/q_y$. In fact, the size of the pulse, i.e., 
the spatial extension of the whole pulse is smaller than $2\pi /q_y$.
This clearly shows that the amplitude equation approach is invalid here. 
Therefore we
believe that the localized pulse observed here is indeed a novel structure 
due to non-adiabatic effect,
which can only be studied using MSHE where variation in all the 
length scales are kept.

According to the above analysis, the long worm can be understood as the 
combined structure of the 1d localized 
state in y direction and the simple harmonic behavior in x direction. 
In the y direction, the amplitude of the worm is maximum at the center
and decays symmetrically as we move away from the center.
The phases on the two sides have effective 
wavevectors $\vec{k_\pm}=(k_x,\pm k_y)$ which are different from the linearly
most unstable modes $\vec{q_\pm}=(q_x,\pm q_y)$. The long worm
expands while moving with the group velocity.

In summary, we have proposed a modified Swift Hohenberg model to count for the
formation of localized worm state seen experimentally in electroconvection
experiment. For a broad parameter range, we have found a solution of the MSHE
which is localized in one direction and extended in the other.
The structure of our solution resembles the experimental findings closely.
The localization of the solution is further understood by simplifying
the original two dimensional equation to 1D. A novel localized stationary
pulse state is discovered in the 1D study which explains the worm structure.

To understand the experiments fully, we need to understand the dynamics of the
worm in detail. For our current model, once a worm is formed, it will persist
until it interacts with the boundary or another worm, while in the experiment,
it seems that it can dissappear by itself. Such a phenomena could be due to
the strong external fluctuation or certain internal 
instability,
which depends on the details of nonlinear terms. Another interesting issue is
the interaction between worms, which also depends crucially on the form of 
the nonlinear terms. The MSHE can also be used to study the spatially extended
STC state observed for larger electrical conductivity and the transition 
between the worm state and the STC state.  

The author would like to acknowledge useful discussion with Drs. M. C. Cross
, H. Riecke and M. Dennin.

\begin{figure}
\caption{(a)Grey scale plot of $Re(\phi(x,y))$ showing localized 
worm structures. (b)The profile of the 2D pattern in (a) at x=45: 
$Re(\phi(45,y))$ versus y, the localized nature of the worm state is obvious.}
\label{fig1}

\caption{Space time plot of $Re(\psi(x,t))+t/4$ versus x for time difference
$dt=4$}
\label{fig2}

\caption{(a)The amplitude of the 1D pulse $A(y)$ versus $y$; (b)The stationary 
part of the phase of the 1D pulse $\alpha_0 (y)$ versus $y$.}
\label{fig3}
\end{figure} 

\end{multicols}
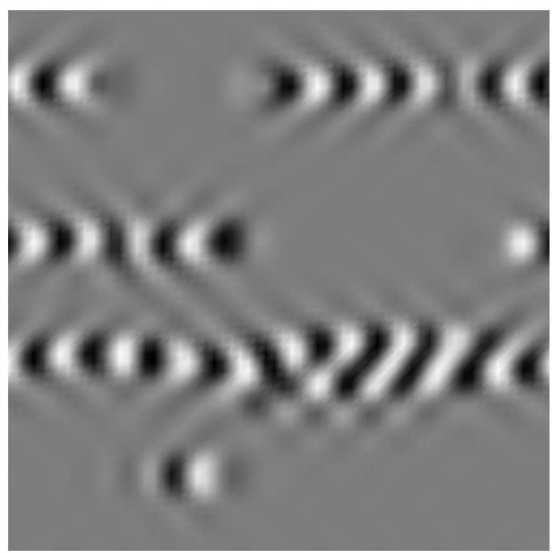
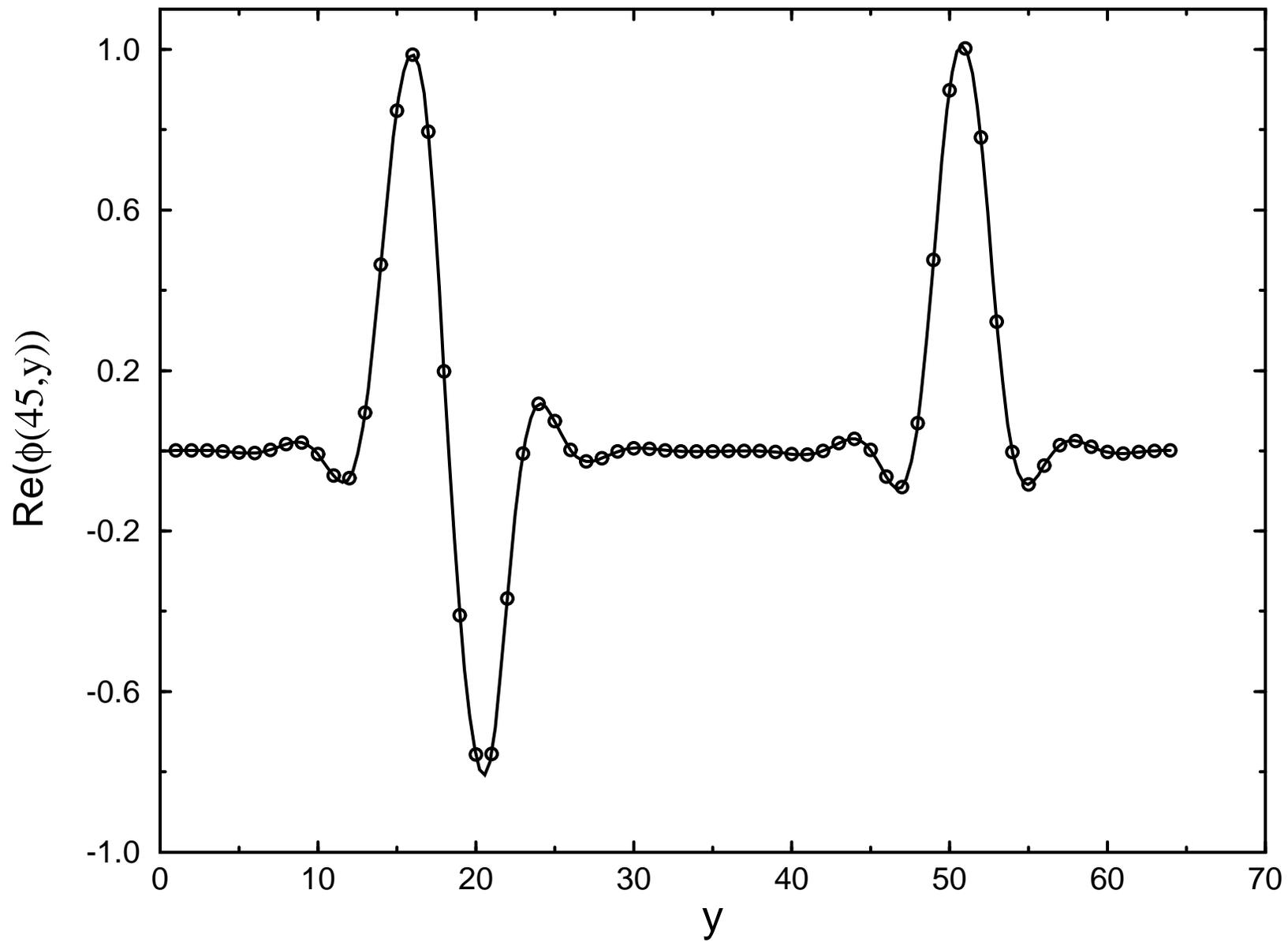
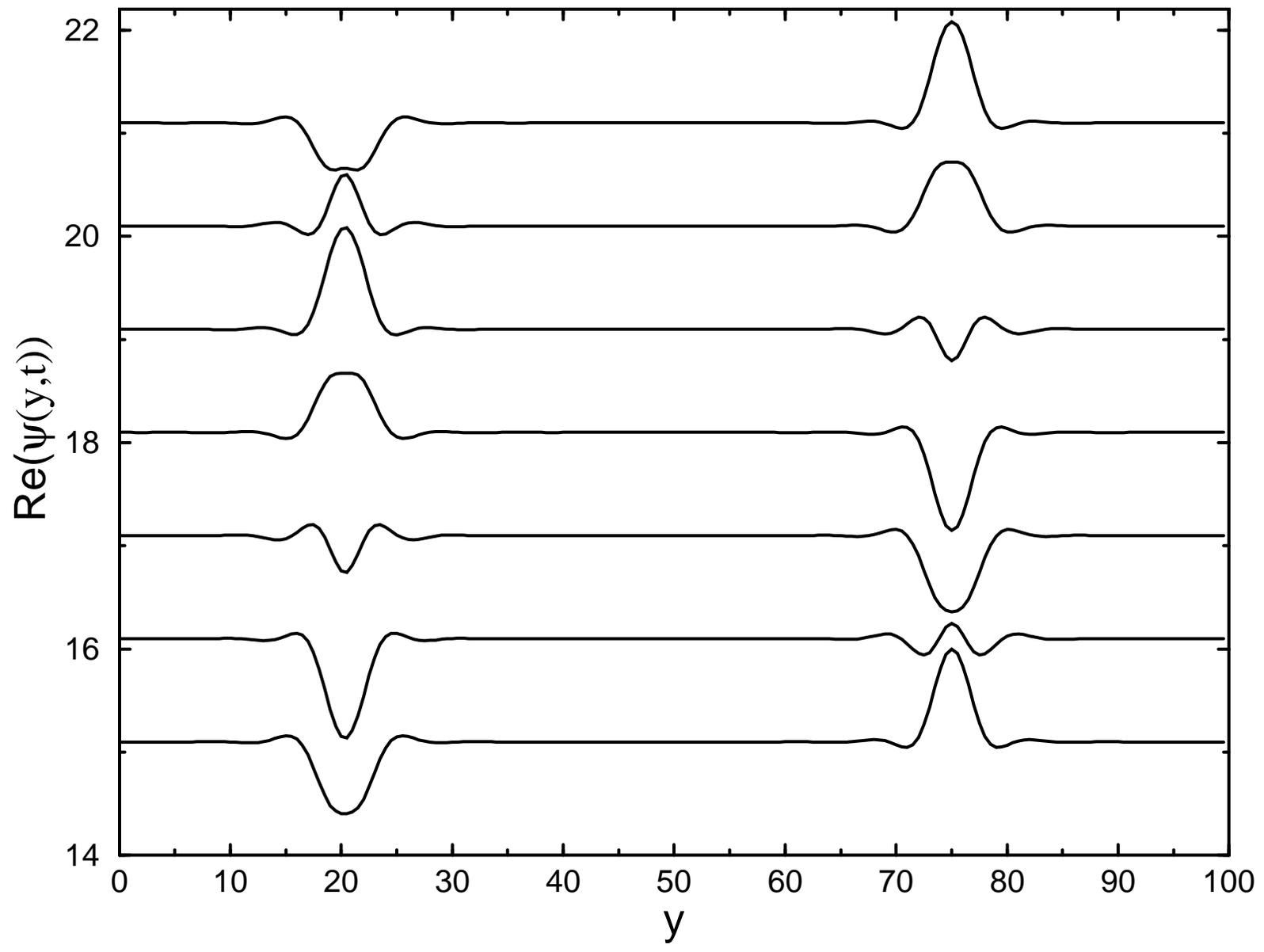
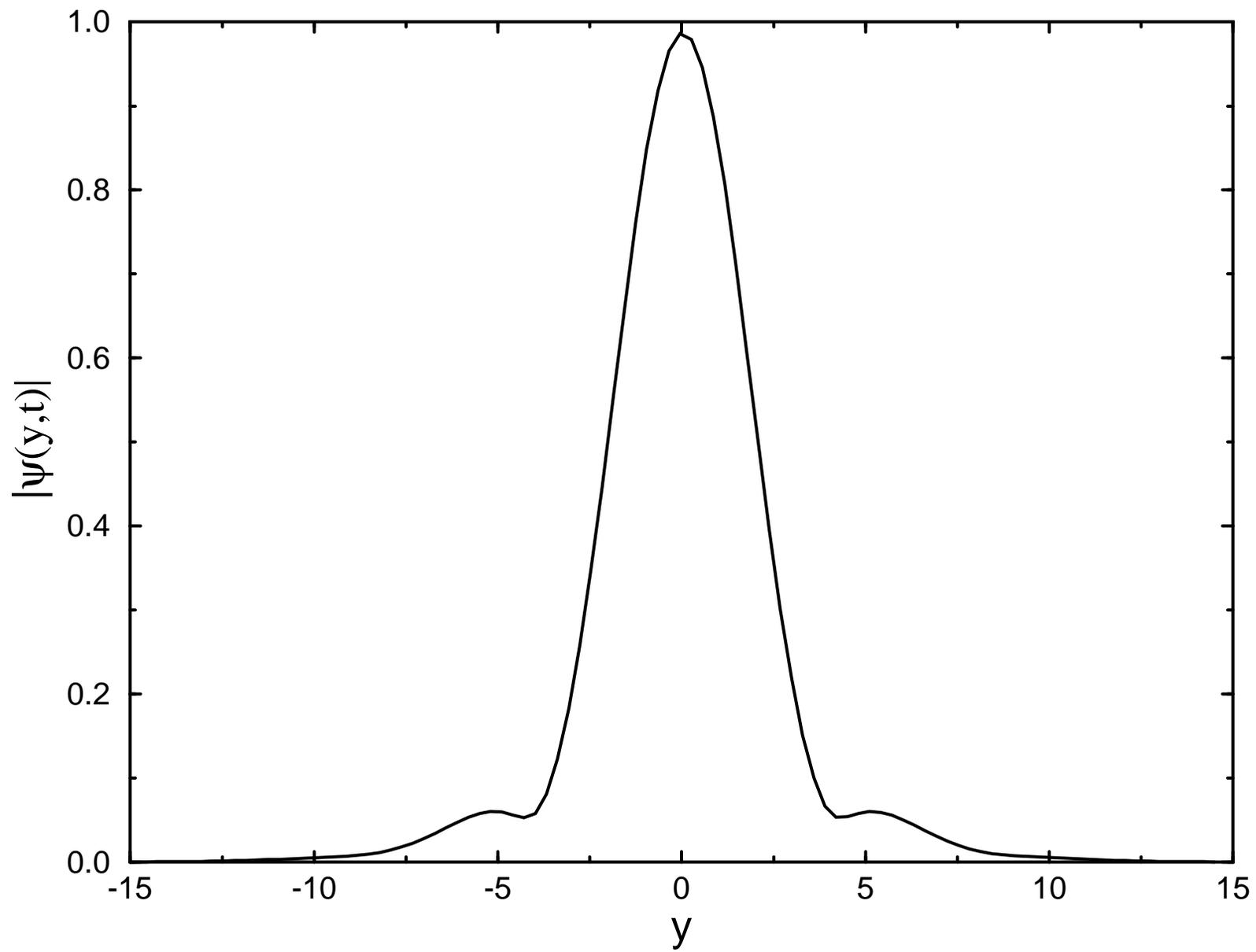
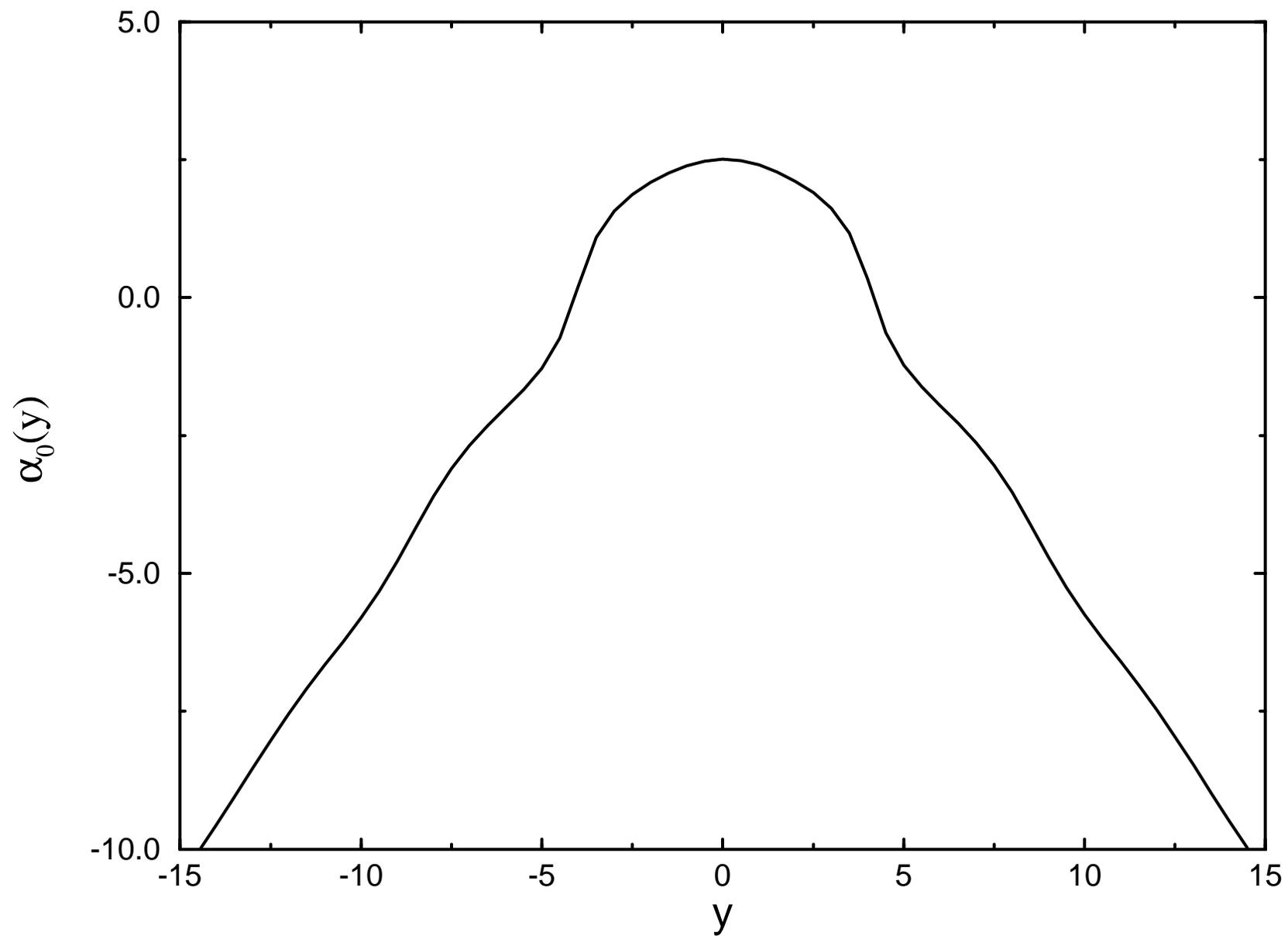
\end{document}